\documentclass[prb,twocolumn,preprintnumbers,amsmath,amssymb,superscriptaddress,showpacs]{revtex4-1}
\usepackage{graphicx}
\usepackage{dcolumn}
\usepackage{bm}
\usepackage{psfrag}

\newcommand{\qvec}{{\bf q}} 
\newcommand{\kvec}{{\bf k}}

\usepackage[colorinlistoftodos,prependcaption,textwidth=1cm,textsize=footnotesize,color=lime]{todonotes}

\usepackage{cleveref}

\begin{document}

\title{Phase separation and proximity effects in itinerant ferromagnet - superconductor heterostructures} 

\author{C. Martens}
\affiliation{Institut F\"ur Physik, BTU Cottbus--Senftenberg,
  PBox 101344, 03013 Cottbus, Germany}
\author{A. Bill}
\email[]{andreas.bill@csulb.edu}
\affiliation{Department of Physics \& Astronomy, California State University Long Beach, Long Beach, CA 90840, USA}
\author{G. Seibold}
\email[]{goetz.seibold@tu-cottbus.de}
\affiliation{Institut F\"ur Physik, BTU Cottbus--Senftenberg, PBox 101344,
  03013 Cottbus, Germany}
\date{\today}

\begin{abstract}
Heterostructures made of itinerant ferromagnets and superconductors are studied. In contrast to most previous models,  ferromagnetism is not enforced as an external Zeeman field but induced in a correlated single-band model (CSBM) that displays itinerant ferromagnetism as a mean-field ground state. This allows us to investigate the influence of an adjacent superconducting layer on the properties of the ferromagnet in a self-consistent Bogoliubov-de Gennes approach. The CSBM  displays a variety features not present in the
Zeeman exchange model that influence the behavior of order parameters close to the interface, as e.g. phase
  separation and the competition between magnetism and superconducting orders. 
\end{abstract}


\maketitle

\section{Introduction}

Superconductivity and ferromagnetism were long believed to be mutually 
exclusive phases because in a conventional superconductor the spin part of 
the wave-function is a singlet which is easily broken by the strong 
magnetization of the ferromagnet. However, there are situations where the 
coexistence is possible as in an itinerant ferromagnet where the spin-up and 
spin-down bands are split  by an effective exchange field. 
Following the idea of Fulde, Ferrell, Larkin and Ovchinnikov  \cite{fflo} 
one can construct a Cooper pair which is composed of two electrons with
opposite spins and momenta,but acquires a finite total momentum due to the exchange field splitting
and resulting different Fermi momenta.
Another possibility is the formation of 'spin triplet' Cooper pairs where 
the associated orbital part is antisymmetric either in the exchange of the 
electrons  position or the time coordinates, the latter being called 
'odd frequency  pairing'.\cite{berez75}

An ideal playground to study these possibilities and many related   questions 
are heterostructures  composed of superconducting and magnetic layers. 
Besides their relevance  for studying fundamental properties of the interplay 
between  superconductivity and magnetism such systems have also a strong 
relevance for applications in spintronics due for example to the possibility of 
acting as spin-polarized current sources.\cite{eschrig}   

The attractive feature of nanohybrid structures is that the two phases   
are spatially separated and interact solely via the proximity effect.  
In a S/F/S junction (S being a superconductor, F a ferromagnet)  Cooper pairs 
tunnel through the S/F interfaces and, for a thin enough ferromagnetic layer, 
may realize a Josephson junction.    

First studies on such systems where conducted by Buzdin and collaborators 
\cite{buzdin82} who predicted for example the existence of  $\pi$-junctions. 
Here the oscillatory behavior of the superconducting state wave function 
leaking into the ferromagnet may lead to a reversal of the Josephson current 
(the $\pi-$state) with respect to the ordinary Josephson effect 
(correspondingly called the $0-$state) if the thickness of the ferromagnet
is chosen such that there is a sign change of the wave function on
either side of 
the junction.

In a series of papers Bergeret, Efetov  and Volkov,\cite{bergeret00} predicted that under certain conditions, such as for example 
the presence of inhomogeneities in the magnetization, a triplet component with non-zero spin projection along the quantization axis,$m = \pm 1$ of the pair amplitude might arise, even though the Cooper pairs generated in the superconducting parts of the junction are singlets. A similar conclusion was reached in Ref.~\onlinecite{kadigrobovEPL01}. Singlet and spin-zero triplet projection pairs ($m = 0$)  
entering a ferromagnet are subject to the pair breaking effect and are thus 
short-range components decaying exponentially over a few nanometers, except at every change of direction of the magnetization where they are regenerated through the cascade effect.\cite{bakerEPL14,bakerPRB16}
In contrast, non-zero spin projection   triplet components are comparatively 
long range because unaffected by   magnetism and have in fact been observed in 
a number of experiments  (see e.g. Refs.~\onlinecite{keizer,sosnin,robinson,anwarPRB10,birgegroup,zhuPRL10,leksinPRL12,wenEPL14}).

The creation of such long-ranged triplet components out of a singlet   
superconductor requires a rotation of the magnetization which can be either  
realized via magnetic multilayers, \cite{birgegroup,robinson} conical magnets as 
Ho \cite{sosnin,robinson} or Heusler alloys. \cite{sprung}  On the theoretical side 
the conditions for which a long range triplet component of the order parameter 
can be  observed in superconductor -- magnetic nanojunctions has been worked 
out in some detail in the diffusive regime (i.e. on the basis of Eilenberger 
and/or Usadel equations; see {\it e.g.} Refs.\onlinecite{bergeret00,bergeret01,belzig,bakerNJP14,bakerEPL14,derojasJSNM12,bakerPRB16,bakerX18} and references therein) 
and in the clean limit within the Bogoliubov-de Gennes (BdG) approach 
(see {\it e.g.} Refs. \onlinecite{halter04,halter05,halter07,halter08,halter12,leadbeater99,zhu00,eschrigJLTP07,fritsch14,degennes,halter02,halter02a}).

More recently other sources for the generation of a long-ranged
triplet component have been analyzed, in particular it has been shown
that the physical mechanism of the singlet- triplet conversion can be
linked to the local SU(2) invariance of magnetized systems with spin-orbit
interaction.\cite{berg13}

In most of the studies of S/F junctions a strong ferromagnet is considered
with a N\'{e}el temperature much larger than $T_c$ so that the influence
of superconductivity on the magnetism is negligible. However, it
was suggested \cite{buzdin88,bergeret00} that under certain conditions for a ferromagnetic film
attached to a S it is favorable to be in a 'cryptoferromagnetic state', \cite{anderson59} i.e. a segregation of the ferromagnet into small-size domains, smaller than the superconducting coherence length.

In this paper we aim to investigate the influence of S on the magnetism
in a S/F system in the clean limit on the basis of the Bogoliubov-de Gennes
approach. We will describe both the superconductor
and the ferromagnet on equal footing i.e. the latter is not implemented
by an external exchange field but also
described by an itinerant model with ferromagnetic exchange $J$ between the
charge carriers (cf. also Refs.~\onlinecite{cuoco09,annunz09}).
Our model is therefore implemented on a finite lattice in contrast
to previous continuum studies of proximity effects within the BdG
approach.\cite{halter02,halter02a,halter04,halter05,halter07,halter08,halter12}

The approach has the advantage that it allows for the study of both proximity effects,
S order parameter in the F and ferromagnetic order parameter in the S as
a function of the exchange $J$, thus covering the range from soft
to hard F's. Moreover, we will investigate the inverse proximity
effect, in particular the weakening of the ferromagnetic order in the
F close to the interface.

Previous studies of heterostructures on discrete lattices
within BdG have addressed the subgap conductance at FS interfaces,\cite{leadbeater99} the decay of d-wave correlations inside the
F within an extended Hubbard model,\cite{zhu00}
or the generation of parallel spin triplets in conical magnetizations such as Holmium.\cite{fritsch14}

Our paper is organized as follows. The next section outlines the
system, model and approximations of our investigations.
The resulting charge density, magnetization and order parameter profiles
together with the corresponding spectral properties are discussed in
Secs. \ref{sec:resF} and \ref{sec:resH}. Finally, the findings are summarized in Sec. \ref{sec:conc}.

\section{Model and Formalism}

\begin{figure}[htb]
\includegraphics[width=8cm,clip=true]{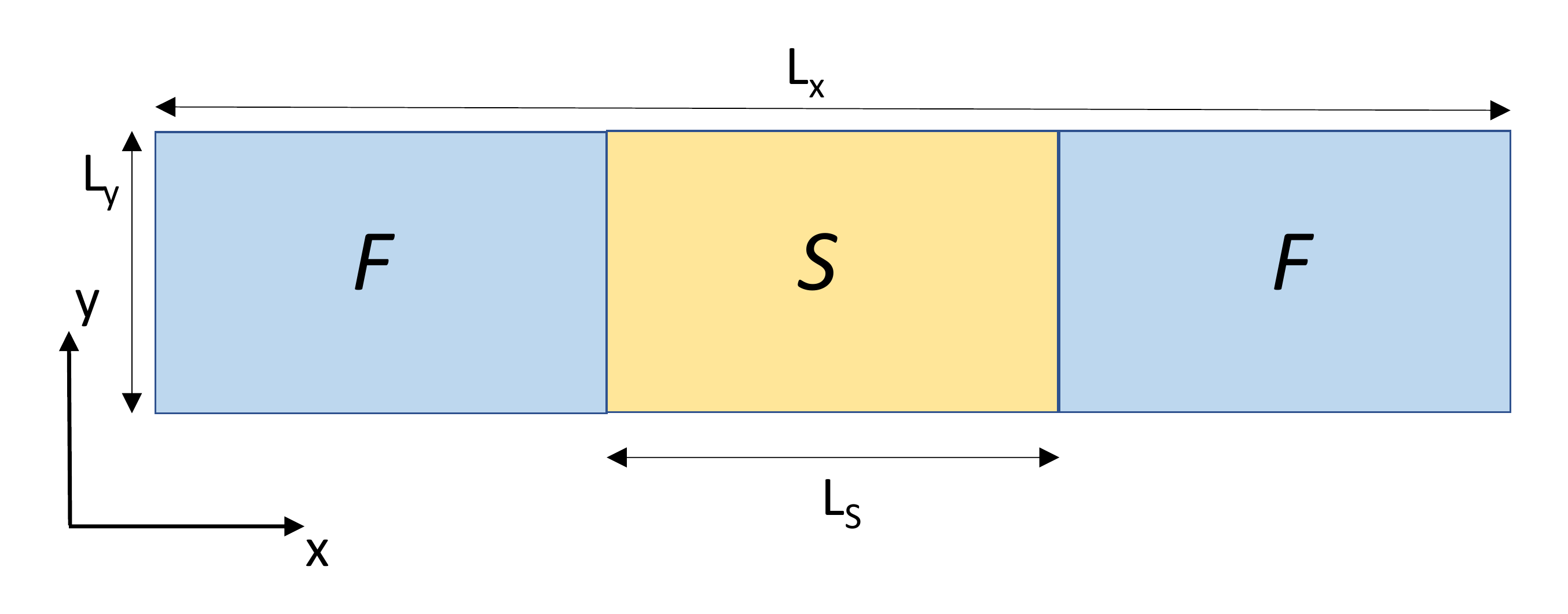}
\caption{\label{fig0}
Sketch of the SF bilayer proximity system with periodic
  boundary conditions along x- and y-directions. $L_{x,y}=N_{x,y}a$ where $a$
  denotes the lattice constant which in the following is set to $a\equiv 1$.}
\end{figure}

We consider a two-dimensional superconducting system sandwiched between two ferromagnetic
layers as shown in Fig.~\ref{fig0}. We use periodic boundary conditions
along x- and y-directions so that the S system is actually connected
to the same ferromagnet on the left and right. Translational invariance
is assumed along the y-direction. The calculations are done at zero temperature.

Our model hamiltonian  reads

\begin{equation}\label{eq:hhi}
  H=H_0 + H_U + H_{int},
\end{equation}
where
\begin{eqnarray}\label{h0}
  H_0&=&\sum_{ij,\sigma}t_{ij}c_{i,\sigma}^\dagger c_{j,\sigma} +\sum_{i,\sigma} (V^{loc}_i-\mu) c_{i,\sigma}^\dagger c_{i,\sigma} \nonumber \\
  && + \sum_{i,\sigma} h_{i,z} \sigma c_{i,\sigma}^\dagger c_{i,\sigma}
\end{eqnarray}
is the single-particle part composed of the kinetic energy and a local term. The latter consists of the chemical potential $\mu$ and a local energy $V^{loc}_i$ which is implemented in order to account for the local orbital energy but also  to tune the charge density in the S layer. We consider electrons on a $N_x\times N_y$ square lattice and $c_{i,\sigma}^{(\dagger)}$ annihilates (creates) a particle with spin $\sigma=\pm 1$ at site $\mathbf{R}_i$; the site index is a two-dimensional vector, $i = (i_x,i_y)$.
The hopping matrix element of the first term is set to $t_{ij}= - t < 0$ for nearest neighbors and $t_{ij}=0$ otherwise.
We also consider a constant Zeeman term $\sim h_{i,z}$ to describe the local magnetization at site $i$. This component of the Hamiltonian is used below to compare the results for the itinerant F system studied in this paper, with previous investigations where F was modeled with an external exchange field $\mathbf{h}$.

The second part of Eq.~\eqref{eq:hhi},
\begin{equation}\label{eq:hu}
  H_U=\sum_i U_i n_{i,\uparrow} n_{i,\downarrow},
\end{equation}
is a local interaction with $n_{i,\sigma}=c_{i,\sigma}^\dagger c_{i,\sigma}$.
Within the superconducting layers we take $U_i\equiv U_S < 0$, i.e. an attractive Hubbard interaction, in order to support
singlet superconductivity.
In the ferromagnetic layers $U_i$ arises as the first term in an expansion of the long-range Coulomb interaction in Wannier functions and therefore $U_i \equiv U_F \ge 0$ in these regions.

Finally, we model the magnetic interaction in the ferromagnetic layer as
\begin{equation}\label{eq:hint}
  H_{int}= -\sum_{\langle ij\rangle}J_{ij} \left\lbrack {\bf s}_i{\bf s}_j + n_i n_j\right\rbrack,
\end{equation}
where $n_i=n_{i,\uparrow}+n_{i,\downarrow}$, $s_i^\alpha=\sum_{\sigma\sigma'}
c_{i,\sigma}^\dagger\tau^{\alpha}_{\sigma\sigma'}c_{i,\sigma'}$ with the Pauli matrices $\tau^{\alpha}$, and $\langle ij\rangle$ limits the summations to nearest neighbor sites. The interaction Eq. \eqref{eq:hint} originates from the nearest neighbor Coulomb interaction terms $J= \langle ij | 1/r | ji \rangle = \langle ii | 1/r | jj \rangle > 0$ and has been derived in
Refs.~\onlinecite{hirsch89,hirsch96,hirsch99} from the expansion of the long-range Coulomb interaction in Wannier functions.
In the expansion we have neglected a density dependent 'correlated hopping' term arising from Coulomb matrix contributions
$~ \langle ii|1/r|ij\rangle$ between sites $\mathbf{R}_i$ and $\mathbf{R}_j$, which has been shown to further stabilize
ferromagnetism.\cite{kollar01,schiller99}
Based on a variational Ansatz it has been argued that the model explains the occurrence of weak metallic ferromagnetism in materials with a partially filled nondegenerate band, as in Sc$_3$In \cite{hirsch91} and metallic hydrogen.\cite{oles92} Note, however, that the model does not capture ferromagnetism in transition metals, where magnetism is usually generated from localized electrons.\cite{kollar01,oles92} Nevertheless, as shown below, the theory contains many features that are also present in double exchange models.
No specific material is considered here and Eq. \eqref{eq:hint} is chosen as a convenient way to describe itinerant ferromagnetism. 

In the heterostructure of Fig.~\ref{fig0}, the ferromagnetic coupling $J_{ij}$ is only finite in the F layer. Furthermore, magnetism and charge density solely vary along the $x$ direction and are constant along the $y$ direction. Therefore, Eq. \eqref{eq:hint} is rewritten in the form
  \begin{eqnarray}
    H_{int}&=&-\sum_{i=(i_x,i_y)}J^x_{i} \left\lbrack {\bf s}_i{\bf s}_{i+\hat{x}} +
    n_i n_{i+\hat{x}}\right\rbrack \label{eq:hint2} \\
&& - \sum_{i=(i_x,i_y)}J^y_{i} \left\lbrack {\bf s}_i{\bf s}_{i+\hat{y}} +
    n_i n_{i+\hat{y}}\right\rbrack\nonumber.
  \end{eqnarray}
In the calculations of later sections it is assumed that $J_i^x = J_i^y = J$ is a constant within the F layer since a moderate anisotropy did not seem to affect the results significantly.

Since the system is translationally invariant along the $y$ direction we perform the corresponding Fourier transform
\begin{equation}\label{eq:trafky}
  c_{i,\sigma}=\frac{1}{\sqrt{N_y}}\sum_{i_x} c_{i_x,\sigma}(k_y) \exp(-ik_y i_y),
\end{equation}
so that the kinetic term Eq. \eqref{h0}
reads ($t>0$)
\begin{eqnarray}\label{h01}
  H_0&=&-t \sum_{i_x,k_y,\sigma}\left\lbrack
  c_{i_x,\sigma}^\dagger(k_y) c_{i_x+1,\sigma}(k_y) + h.c.\right\rbrack \\
  &+& \sum_{i_x,k_y,\sigma}\left\lbrack -2t \cos(k_y)- \mu\right\rbrack
  c_{i_x,\sigma}^\dagger(k_y) c_{i_x,\sigma}(k_y). \nonumber
\end{eqnarray}

We apply the transformation, Eq.~\eqref{eq:trafky}, to the interaction terms, Eqs.~(\ref{eq:hu}-\ref{eq:hint2}). We then approximate these terms in mean-field. This includes the anomalous singlet (Gor'kov) correlations $f_0(i)=
\left\langle c_{i,\downarrow}c_{i,\uparrow}\right\rangle$ that are induced in the S regions where $U_i<0$ but leak into the F layer due to the proximity effect.
The problem then can be diagonalized by means of the Bogoliubov-Valatin transformation 
\begin{equation}
    c_{i_x,\sigma}(k_y)=\sum_p \left\lbrack u_{i_x,\sigma}(p,k_y)\gamma_{p,k_y}
      -\sigma v^*_{i_x,\sigma}(p,k_y)\gamma^\dagger_{p,k_y}\right\rbrack ,
\end{equation}
and the integer $p$ labels the eigenvalue.
Introducing the basis vector
$\vec{\Psi}_n(p,k_y)=\lbrack u_{n,\uparrow}(p,k_y),u_{n,\downarrow}(p,k_y),
v_{n,\uparrow}(p,k_y),v_{n,\downarrow}(p,k_y)\rbrack$
one has to solve the following eigenvalue problem for each value of $k_y$

\begin{equation}\label{eigenvalueproblem}
  \underline{\underline{H}}_{ij}(k_y)\vec{\Psi}_j(p,k_y)=\varepsilon_p(k_y)
  \vec{\Psi}_i(p,k_y),
\end{equation}

where the hamiltonian is composed of a local and an intersite part
\begin{equation}\label{eq:hammat}
  \underline{\underline{H}}_{i_xj_x}(k_y)=\underline{\underline{T}}_{i_x,j_x}(k_y) \left( \delta_{j_x, i_x+1} + \delta_{j_x,i_x-1} \right)
  +\underline{\underline{V}}_{i_x}(k_y)\delta_{j_x,i_x}\,.
\end{equation}
The explicit structure of these operators is given in appendix \ref{apa}.

\section{Ferromagnetic system}\label{sec:resF}

Before presenting the results for the SF heterostructure we
briefly discuss the homogeneous magnetic system, i.e. $U_i\equiv U_F>0$
in Eq. \eqref{eq:hu} and $J^x_i=J^y_i\equiv J$ in Eq. \eqref{eq:hint2}. This case is instructive for the later analysis of the competition between F and S in the interface regions. A more extensive discussion of the magnetic system can be found in Refs. \onlinecite{hirsch89,hirsch96,hirsch99,oles92}.
Details of the calculations are provided in appendix \ref{apb}.

The chosen values of parameters, $U_F/t$ and $J/t$, are generic but describe realistic systems. For example, $U_F$ is up to the order of the bandwidth, while $J$ is typically less than $U_F$.\cite{hirsch89} Calculations are performed on a lattice with $N_x\times N_y = 420 \times 420$ sites.

The model already displays rich physics for a single magnetic layer.
Depending on the parameter values the system is paramagnetic, ferromagnetic,
antiferromagnetic or shows electronic phase separation.\cite{hirsch89,hirsch96,hirsch99,oles92}

\begin{figure}[hhh]
\includegraphics[width=9.5cm,clip=true]{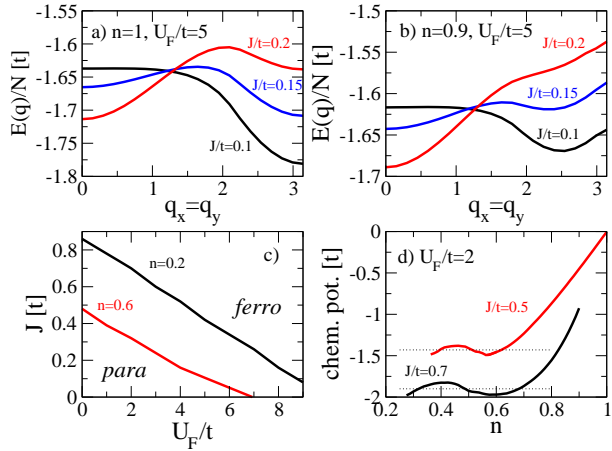}
\caption{\label{fig0a}
Ground state energy vs. spin-spiral modulation vector $\mathbf{q}$ $\left(S_i = S_0 \exp{\left(i \mathbf{q}\cdot \mathbf{R}_i\right)}\right)$ along the diagonal direction of the Brillouin zone at half-filling (a) and density $n=0.9$ (b). Note that for other $U_F/t$ results are qualitatively similar but energy variations decrease with decreasing $U_F/t$. (c) Phase diagram in the $(U_F/t,J/t)$-plane for density $n=0.2$. (d) Chemical potential vs. density. The horizontal dotted lines are determined by the Maxwell construction. The compressibility diverges at the local extrema.}
\end{figure}

Panels (a,b) of Fig. \ref{fig0a} report the ground state energy $E({\bf q})$
  for a spiral modulation $S_i = S_0 \exp{\left(i \mathbf{q}\cdot \mathbf{R}_i\right)}$ as a function of the spiral wave-vector ${\bf q}$ which is taken along
  the diagonal direction ($q_x=q_y$). $S_0$ is a variational parameter.

At half-filling and $U_F/t>0$, $J/t=0$ the system shows antiferromagnetic spin-density wave order [${\bf q}={\bf Q}_{AF}=(\pi,\pi)$] which due to perfect nesting occurs for infinitesimally small values of the repulsive interaction $U_F/t$.\cite{nagaoka} 

Upon increasing the ferromagnetic exchange $J/t>0$ a second energy minimum develops at ${\bf q}=(0,0)$ which above some critical $J/t$ that depends on $U_F/t$ corresponds to the ferromagnetic ground state [cf.~Fig.~\ref{fig0a}(a)]. As can be seen from panel (b) of Fig.~\ref{fig0a} the same holds for doping away from half-filling where for sufficiently large repulsion $U_F/t$ and $J/t=0$ the commensurate AF is replaced by a spiral, but with some incommensurate modulation ${\bf Q}_{spiral}=(q,q)$.
In the regime of small doping ($n \ll 1$) and small $U_F/t$ ($< 1/N(E_F)$)
the system would be a paramagnet for $J/t=0$ and ferromagnetism can be
induced above some critical $J/t$. The corresponding phase diagram is displayed in Fig.~(\ref{fig0a}c) for concentrations $n=0.2$ and $n=0.6$.
Upon increasing $U_F/t$ the transition line approaches the value for the
standard Stoner criterion $U_F=1/N(E_F)$ at $J=0$ ($N(E_F)$ is the density of states at the Fermi energy $E_F$).

In Figure \ref{fig0a}(d) we also demonstrate that the model has an instability region with respect to phase separation which can be deduced from the dependence of the chemical potential $\mu$ on the density $n$. The compressibility $\kappa=\partial n/\partial\mu$ diverges at the local extrema of $\mu(n)$ and becomes negative in between. The phase separation region in $n$ is determined by the Maxwell construction (dotted horizontal line). The two curves in Fig.~\ref{fig0a}d indicate that the phase separation region decreases with decreasing $J/t$. Note that the occurrence of phase separation is {\it not} a peculiar feature of the present model. This phenomenon also appears in double exchange models that are for example used for the description of magnetism in manganites (cf. Ref. \onlinecite{dagotto} and references therein).

\section{Superconducting-magnetic heterostructure}\label{sec:resH}

The results of this section have been obtained on lattices with $120\times 80$ sites and periodic boundary conditions in both directions. In the S region ($40 \le i_x \le 80$) singlet superconductivity is generated with a negative $U_S=-2t$. For this value the coherence length can be estimated as $\xi_S\approx 4$ in units of the lattice spacing, i.e. much smaller than the linear size of the system.

The remaining sites pertain to the F region with local onsite repulsion $U_F>0$ and ferromagnetic exchange interactions $J_i^x\equiv J^x$, $J^y_i\equiv J^y$ and $J^x=J^y=J$. This description of the F layer will be referred to as the correlated ferromagnetic model (CFM). For comparison, we also use an exchange field $h_{z}$ to model the F. The latter is referred to as the exchange field model (EFM).

Since we consider an itinerant F we treat superconductivity and magnetism on equal footing. Hence, we present in this section results for the charge density, the magnetization and pair correlations in both the F and the S.

In addition to the normal proximity effect, two distinct phenomena appear in these hybrid structures: the inverse proximity effect and phase separation. In the first, the S correlations suppress the magnetization inside the F
near the SF interface.
In general, $n_i\neq m_i$ in such situation; the F is nowhere fully polarized.
Moreover, the coupling between magnetic and charge degrees of freedom leads to
a concomitant reduction of the $n$ near the SF interface.
By contrast, when the system undergoes phase separation, the system is fully polarized ($n_i = m_i$) deep in the F and the superconducting state is affected by the itinerant electrons of the F.

\subsubsection{Charge density, magnetization and pair correlations}

\begin{figure}[htb]
\includegraphics[width=8.5cm,clip=true]{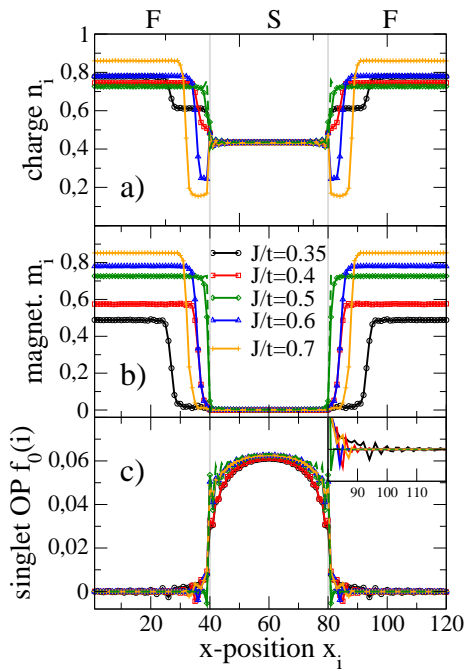}
\caption{\label{fig1}
Charge density (a), magnetization (b) and singlet pair correlations (c) as a function of $x$ and various values of the exchange coupling $J/t$ in
  the ferromagnetic region. For all cases the local potential $V^{loc}$ is adjusted in such a way as to obtain a similar charge density deep in the S layer. The thin dashed green line reports the result within the EFM for $h_{i,z}/t=3$ (see text). 
Parameters: $U_F/t=2$, $U_S/t=-2$, $n = 0.625$.
}
\end{figure}

\begin{figure}[htb]
\includegraphics[width=8.5cm,clip=true]{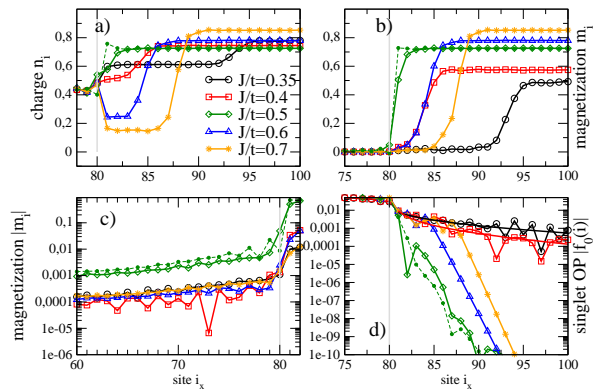}
\caption{\label{fig2}
Close look near the interface of the charge density (a), magnetization (b,c) and singlet pair correlations (d) shown in Fig. \ref{fig1} for various values of the exchange coupling $J/t$ in the F. For all cases the local potential $V^{loc}$ is adjusted as described in Fig.~\ref{fig1}. Shown for comparison as a thin dashed green line is also the pair correlation in the EFM for $h_{i,z}/t = 3$. The S correlations inside the F in panel d have been fitted with Eq.~\eqref{f0} (light red and black lines; see text for fitting parameters). Parameters: $U_F/t=2$, $U_S/t=-2$, $n = 0.625$.
}
\end{figure}

Figure \ref{fig1} reports the charge density $n$ (panel a), magnetization $m$ (panel b), and singlet pair correlations $f_0$ (panel c) in the heterostructure for varying exchange constant $J/t$. Since this parameter also influences the local Hartree-potential in the F layers a change of $J$ alters the charge distribution between the S and F regions. To be able to compare results for different values of $J$ we therefore adjust the local potential $V^{loc}$ in the S regions in such a way that the charge density is the same for all $J/t$-values deep inside the S layer; hence, in panel (a) of Fig.~\ref{fig1} the charge densities $n$ overlap in S for all $J/t$. Since most of the physics occurs close to the interface between the S and the F layer, Fig.~\ref{fig2} zooms into this region to show the behavior of $n_i$, $m_i$ and $f_0(i)$. In addition, the figure also reports the result for the EFM (dashed lines) for $h_{i,z} = 3t$. The value of $h_{z}$ is fixed in such a way that it reproduces the same magnitude of the magnetization as the CFM deep inside the F region for $J/t = 0.5$. 

From Figs. \ref{fig1} and \ref{fig2} one can distinguish three different regimes. For the parameters of the system, these are $J/t \lesssim 0.3$, $0.3\lesssim J/t \lesssim 0.55$ and $J/t\gtrsim 0.55$. At low $J/t$ ($\lesssim 0.3$) ferromagnetism disappears; this paramagnetic regime was discussed in the previous section.
Above this transition  the {\it inverse proximity effect} regime [black open circles and red squares in panels (a-d)] is effective. The S correlations completely suppress the magnetization in F over a significant distance from the interface. This regime is also characterized by a partial depletion of charge density over similar depth in the F, resulting from the coupling between charge and spin.
The third regime is the {\it phase separation regime} found at high values of $J/t \gtrsim 0.55$ (blue triangles and gold stars). Phase separation was found in the homogeneous system of Sec.~\ref{sec:resF} [see Fig.~\ref{fig0a}(d)]. It is characterized by full polarization deep in the F ({\it i.e.} $n_i=m_i$) and a concomitant depletion of the charge density and the magnetization in the F over moderate distance away from the interface. 

We note that for intermediate values, $J/t\approx 0.5$ (green diamonds), the charge and magnetic profiles adjust to reach equal value already within the charge/spin correlation length $\sim 1/k_F$ ($\approx 2-3$ lattice constants) from the SF interface. This steep rise is close to the result of the conventional EFM (dashed green line in Figs.~\ref{fig1},\ref{fig2}), where the transition is driven by the abrupt onset of the magnetization inside the ferromagnetic layer.

Figure \ref{fig2}(c) reveals the behavior of the magnetization in the S. The magnetization decays exponentially with a correlation length (naturally) independent of $J/t$. The overall magnitude is determined by the value of $m_i$ at the interface, which is largest for intermediate values of $J/t$, where $m_i$ is not suppressed by S correlations and phase separation is not relevant (diamond green lines in Fig.~\ref{fig2}).

Figs.~\ref{fig1}c and \ref{fig2}d display the decay of the superconducting order parameter $f_0(i)$ in the F. Panel \ref{fig1}(c) shows the overall behavior of the singlet order parameter for the same range of $J/t$ values while the decay inside the F is detailed in the inset, and in panel (d) of Fig.~\ref{fig2} on a logarithmic scale. The latter figure shows the stark contrast between the inverse proximity ($J/t \lesssim 0.55$) and the phase separated ($J/t > 0.55$) regimes. Both cases can be modeled by the following expression
\begin{eqnarray}
\label{f0}
f_0 \sim \frac{e^{\left(-x/\xi_N\right)}}{x} \cos\left(\frac{x}{\xi_F}\right),
\end{eqnarray}
where $\xi_N= v_F/2\pi T$ and $\xi_F = v_F/2\pi m(x)$ are the paramagnetic and ferromagnetic coherence lengths. A close look at the curves in Fig.~\ref{fig2} shows that the magnetization takes non-zero values from the interface on; for example, for $J/t = 0.35$ the value of $m(x)$ is small but finite already for $80<x\leq 91$. The above expression for $f_0$, Eq.~\eqref{f0}, can be used to obtain an excellent piecewise fit of the numerical data; one divides the space into $x<x_0$ and $x>x_0$ with $x_0 \sim 82$ ($x_0\sim 90$) for $J/t = 0.35$ ($J/t=0.4$). For $J/t = 0.35$ for example, the curve in regions $80< x \leq x_0$ can be fitted with $\xi_N>L_F$ and $\xi_F = 30$, whereas for $x> x_0$ we have $\xi_N = 33$ but $\xi_F = 0.6$. For $x< x_0$ the pair correlation undergoes a smooth exponential decay that is expected of a paramagnet (the magnitude of $m(x)$ is very small in this region), whereas farther away the behavior is characteristic of a homogeneous ferromagnet.
These results are consistent with previous findings for singlet pair correlations in the EFM, in a paramagnet (Ref.~\onlinecite{degennes}) and a homogeneous ferromagnet (see for example Ref.~\onlinecite{buzdin82}). Refs.~\onlinecite{halter02,halter02a} provided a complementary analysis of the decay of $f_0$ within the EFM. Two almost identical length scales were introduced that inversely scale with the polarization of the F and are $\sim (k_{F,\uparrow}-k_{F,\downarrow})^{-1}$. The oscillatory behavior is due to the interference of up- and down excitations in the pair amplitude.

In the opposite case of large exchange coupling $J/t$ (phase separation regime) the pair correlations first follow the  small-$J/t$ behavior up to some distance $x_0$ away from the interface, followed by a much stronger decay deeper inside the F ($x> x_0$). The length $x_0$ is determined by the point where the magnetization reaches full polarization and therefore increases with $J/t$ due to the increasing low density domain inside the F.

\begin{figure}[htb]
\includegraphics[width=8.5cm,clip=true]{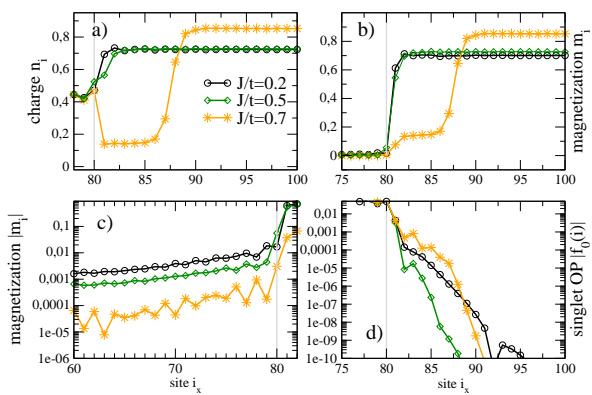}
\caption{\label{fig3}
Charge density (a), magnetization (b,c) and singlet pair correlations (d) close to the interface as in Figs.~\ref{fig1},\ref{fig2} but for a larger value of the Coulomb potential in the F, $U_F/t = 5$. The phase separation regime is substantially extended when compared to Figs.~\ref{fig1}, \ref{fig2}. Both pair correlations in the F and the charge and magnetic configurations in S are rapidly suppressed away from the interface.
 Parameters: $U_F/t=5$, $U_S/t=-2$, $n = 0.625$.}
\end{figure}

The results of Figs.~\ref{fig1}, \ref{fig2} were obtained for $U_F/t = 2$. A larger local correlation stabilizes the F and increases the range of values of $J/t$ for which the F is fully polarized (see Fig.~\ref{fig3}). As a result, even
for values of $J/t\approx 0.2$ close to the onset of ferromagnetism, pair correlations are not able to significantly suppress magnetism; the inverse proximity effect is almost absent. Nevertheless, this reduction of the intermediate regime does not imply a corresponding extension of the phase separation regime to lower values of $J/t$. At $J/t\sim 0.5$ one still observes a behavior similar to the EFM. The phase separation instability for large $J/t \gtrsim 0.5$ persists and the pair amplitude decay is again shifted away from the interface (panel (d) of Fig.~\ref{fig3}). The distance from the interface over which the magnetization is suppressed is about the same than observed in Fig. \ref{fig2}.
As expected, the behavior of the magnetization in the S is unaffected by the change in $U_F/t$ in the F.

\subsubsection{Spectral properties and pair correlations}

The proximity effect is also reflected in spectral properties such as the local density of states (LDOS),
\begin{eqnarray*}
\rho^{loc}(x_i,\omega)&=&\frac{1}{N}\sum_{p,k_y,\sigma}\left\lbrack
|u_{i,\sigma}(p,k_y)|\delta(\omega-\varepsilon_{p}(k_y)) \right. \\
&-& \left. |v_{i,\sigma}(p,k_y)|\delta(\omega+\varepsilon_{p}(k_y))\right\rbrack\,,
\end{eqnarray*}	

which within the BdG formalism and the EFM has been analyzed in Refs. \onlinecite{halter02,halter02a}.

\begin{figure}[htb]
\includegraphics[width=7.5cm,clip=true]{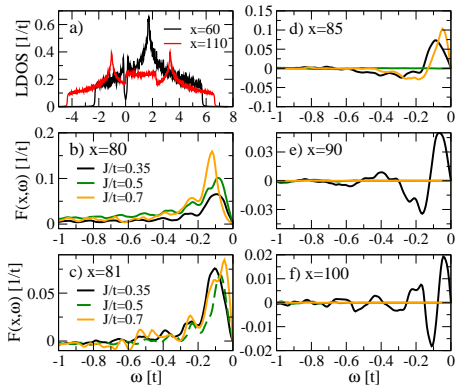}
\caption{\label{fig4}
Local density of states (a) inside the S (black; $x=60$) and the F (red; $x=110$) for $J/t=0.5$. Panels (b-f): imaginary part of the singlet Gor'kov pair correlation $F(i_x,\omega)$ dependence on frequency at specific points in the heterostructure and for various exchange couplings as indicated in the panels. The correlations at $x=80$ are in the S while those for $x\geq 81$ are in the F. Note that the singlet order parameter for $J/t=0.5$ and $x=81$ has opposite sign as compared to the other couplings (cf. panel (c) of Fig.~\ref{fig1}). For better comparison we have therefore multiplied the $J/t=0.5$ in panel (c) by $-1$, symbolized by the dashed line. Parameters: $U_F/t=2$, $U_S/t=-2$, $n = 0.625$.
}
\end{figure}

Fig.~\ref{fig4}a shows the LDOS deep in the S and the F.  Noticeable are the standard BCS coherence peaks at the gap edges in the S (near $\omega = 0$, black curve at $x=60$); a small numerical 'pair-breaking parameter' $\epsilon=0.02t$ has been introduced for numerical reasons which is responsible for the small finite LDOS inside the gap.
The overall structure of the LDOS is otherwise characteristic of a two-dimensional square lattice with its logarithmic van-Hove singularity at the band center.

Deep inside the ferromagnet [red curve at $x=110$ in Fig.~\ref{fig4}(a)] the van Hove singularity is split due to the formation of subbands (peaks near $\omega/t = \pm 3$). 
Note that the apparent "noise" in the data is not due to the lack of precision of the calculation, but are oscillations originating from the discreteness of the lattice.

It is instructive to investigate the dynamical singlet pair correlations (Gor'kov function)
\begin{equation}\label{singc}
f_0(i_x,t)=\frac{1}{2}\left\lbrack \langle c_{ix,\uparrow}(t)c_{ix,\downarrow}(0)\rangle - \langle c_{ix,\downarrow}(t)c_{ix,\uparrow}(0)\rangle\right\rbrack
\end{equation}
inside the ferromagnet for different exchange parameters $J/t$.
The imaginary part of the Fourier transform reads
\begin{eqnarray}\label{eq:fx}
  F(i_x,\omega)&=&\mbox{Im}
\int_{-\infty}^\infty\! dt\,\, \mbox{e}^{i\omega t}f_0(i_x,t)\\
  &=&\!\!\pi\sum_{p,k_y}\left\lbrack u_{i,\uparrow}(p,k_y)
  v^*_{i,\downarrow}(p,k_y)+u_{i,\downarrow}(p,k_y)v^*_{i,\uparrow}(p,k_y)\right\rbrack \nonumber\\
  &\times&\langle \gamma_{p,k_y} \gamma^\dagger_{p,k_y}\rangle \delta(\omega+\varepsilon_{p}(k_y)) \nonumber \\
  &-&\!\! \pi\sum_{p,k_y}\left\lbrack u_{i,\uparrow}(p,k_y)
  v^*_{i,\downarrow}(p,k_y)+u_{i,\downarrow}(p,k_y)v^*_{i,\uparrow}(p,k_y)\right\rbrack \nonumber\\
  &\times& 
  \langle \gamma^\dagger_{p,k_y} \gamma_{p,k_y}\rangle \delta(\omega-\varepsilon_{p}(k_y))\nonumber.
\end{eqnarray}

\begin{figure}[htb]
\includegraphics[width=8.2cm,clip=true]{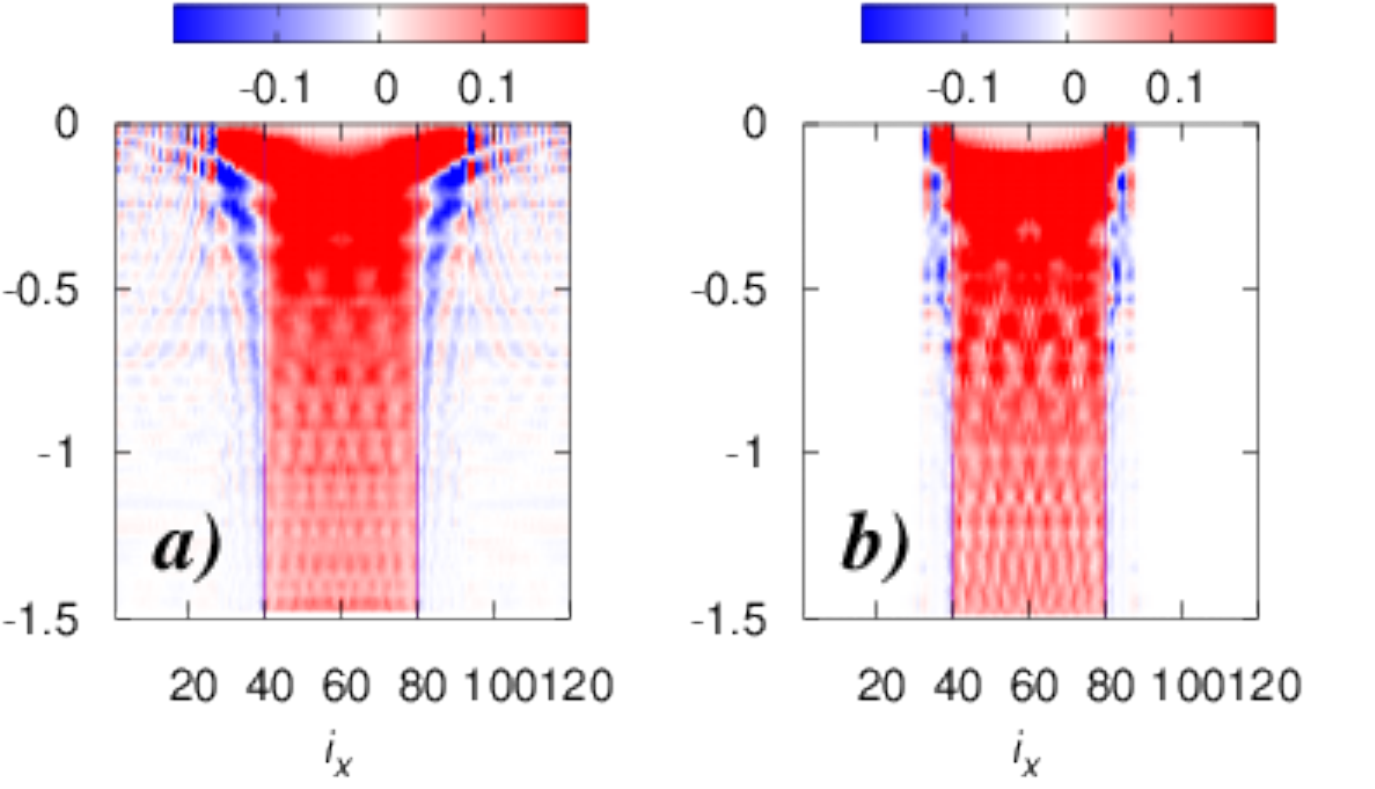}
\caption{\label{fig5}
Position and frequency dependent singlet correlations, Eq. \eqref{eq:fx}, for (a) $J/t=0.35$ and (b) $J/t=0.7$. Pair correlations oscillate both in space and frequency and extend deeper into the F for small $J/t$. Parameters $U_S/t=-2$, $U_F/t=2$ $n = 0.625$.
}
\end{figure}

Figure \ref{fig5} shows the position and frequency dependence of the singlet correlations, Eq. \eqref{eq:fx}, as an intensity plot which visualizes the decay of the pair correlations inside the ferromagnet. Note that the Gor'kov functions are asymmetric in $\omega$, $f_0(-\omega) = -f_0(\omega)$, and in a clean S system show a $\pm 1/\sqrt{\omega-\Delta}$ singularity at the gap edges [$\Delta(x) = f_0(i_x,t=0)$ being the superconducting gap]. Only the $\omega<0$ part of $F(i_x,\omega)$ is shown in Figs.~\ref{fig4} and \ref{fig5}.

A main difference between the inverse proximity regime and the phase separated regime is immediately apparent at small $\omega/t$ when comparing Figs.~\ref{fig5}(a) and \ref{fig5}(b): pair correlations extend deep into the F for small $J/t$ [Fig.~\ref{fig5}(a)]. The region in the F where these pair correlations are present also shrinks with increasing $\omega/t$. In the region where the magnetization is suppressed, either because of the inverse proximity effect (small $J/t$) or phase separation (large $J/t$) the low energy $F(i_x,\omega)$ continuously extends from the S region into the F. This is clearly visible in Fig.~\ref{fig5}(a) at low $\omega/t$ where the intensity plot shows pronounced pair correlations (indicated by solid red color) close to the interface. By contrast, the Gor'kov function starts oscillating within a partially polarized region of the ferromagnet as indicated by the alternating red-blue pattern of Fig.~\ref{fig5}(a). Similarly, oscillations of $F(i_x,\omega)$ are seen at fixed position $i_x$ as a function of $\omega/t$ and we now analyze this $\omega$-dependence of $F(i_x,\omega)$ in more detail at various distances from the interface in the F, shown in panels (b-f) of Fig.~\ref{fig4}.

The singlet pair correlations $F(i_x,\omega)$ shown in panel (b) of Fig.~\ref{fig4} are calculated at the interface, on the S side, $x=80$. The peak position, indicating the size of the S gap is largest (and sharpest) for $J/t=0.7$ in agreement with the larger singlet order parameter (cf. panel (c) of Fig.~\ref{fig1}).

On the other side of the interface, in the F ($x=81$, panel (c) of Fig.~\ref{fig4}) one still observes a sizable gap for all couplings, which however, is now largest for $J/t=0.35$ where the magnetization is suppressed close to the interface. Further away from the interface, at $x=85$, panel (d) shows that for $J/t=0.5$ the
magnetic system is already completely polarized, and the pair correlations are suppressed on the scale of the plot. Interestingly, at this same location, the phase separated solution ($J/t=0.7$) and the inverse proximity solution ($J/t=0. 35$) still reveal a low energy peak and thus the occurrence of a proximity induced gap. Moreover, at larger energies ($\omega/t \sim -0.3$) a second broad peak appears with opposite sign. For $J/t=0.35$ this peak turns out to be related to the onset of frequency oscillation in $F(i_x,\omega)$ observed further away from the interface, as seen in panels (e) and (f).
In the diffusive limit and within the EFM \cite{buzdin82,bergeret00} it was shown that the occurrence of
such oscillations in frequency is a direct consequence of the exchange field $h$. Similarly, in the BdG approach they arise from the superposition of the different excitations in the spin-up and down bands and thus disappear when the system is completely polarized. This explains why in panel (e) and (f), at $x=90, 100$, only pair correlations in the inverse proximity regime ($J/t=0.35$) are finite.
Note that the frequency integration of $F(i_x,\omega)$ yields the local (equal time) Gor'kov function $\Delta(x) = f_0(i_x,t=0)$ at $x$ which vanishes in the F region and thus requires a cancellation of the finite contributions
to $F(i_x,\omega)$.

Finally, we note in panels (d-f) that there are smaller oscillations, both in magnitude and frequency, superposed to the large oscillation of $F(i_x,\omega)$ just mentioned. These are the same small oscillations found in all curves of panels (b,c) and are related to the discrete spatial lattice.

We used everywhere $m$ instead of $S_z$ so i changed it in the following paragraph and removed the sentence that refers to the $z$-axis choice.
The results of this section \ref{sec:resH} were obtained for singlet pair correlations generated in the S and leaking into the F. A similar behavior is also found for the $m=0$ triplet correlations brought about in the F region. In the SF hybrid structure with periodic boundary conditions studied here, there are no $m = \pm 1$ triplet components since the magnetic inhomogeneities are found in the magnitude of the magnetization while its orientation is fixed. This contrasts the magnetic and superconducting inhomogeneities discussed here from those of previous work.\cite{bakerNJP14,bakerEPL14,derojasJSNM12,bakerPRB16,bakerX18}

\section{Conclusion}\label{sec:conc}

We have analyzed proximity effects in a FS heterostructure in which
the F is described within an extended Hubbard-type model where
the ferromagnetic exchange arises from intersite contributions of
the Coulomb interaction. Such a description allows the self-consistent
treatment of both superconducting and magnetic order parameters which
gives rise to features not present in approaches where the exchange
field is fixed inside the F. In particular, we have found that
for small exchange interactions and onsite correlations the magnetization
close to the interface may be suppressed by the S correlations which
significantly alters the decay of the pair correlations inside the F.
Similar for large exchange couplings the system shows an instability
towards phase separation which is also realized close to the
interface with a concomitant suppression of the magnetization.
As a consequence the S correlations extend far inside the phase
separated region and only get suppressed when the magnetization recovers
at some distance from the interface.
The correlations leak much deeper than for the effective field model; this may significantly affect the Josephson current and is being investigated.
Such an instability towards phase separation is also inherent in
double-exchange models for ferromagnetism which are usually considered
to be appropriate for transition metals \cite{dagotto}.
Therefore, we expect that these aspects of our results are
also valid in such systems and are planned in future investigations.
In the present paper we have restricted ourselves to collinear magnetic structures. However, microscopic magnetic models usually  show also more complex
magnetic structures in some part of the phase diagram, as for example the
antiferromagnetic spirals in panel (b) of Fig.~\ref{fig0a} for
$J/t=0.1$. A heterostructure where the ferromagnet displays a spiral
rotation would also induce $m=1$ triplet components inside the
F. As a result, we expect that the pair correlations inside the F have a pronounced influence on the periodicity of such magnetic structures. Work in this direction is in progress.

\section{Acknowledgements}
CM and GS thank the DAAD for financial support. AB gratefully acknowledges funding provided by the National Science Foundation (DMR-1309341).

\appendix
\section{} \label{apa}
The matrices defined in Eq. \eqref{eq:hammat} are given by
\begin{widetext}

\begin{equation}
  \underline{\underline{T}}_{i,i+x}(k_y)=\left(
  \begin{array}{cccc}
    -t+2J_i^x(k^x_{i,\uparrow}+k^x_{i,\downarrow})^* &  0 & 0 & 2J_i^x p^x(i) \\
    0 & -t+2J_i^x(k^x_{i,\uparrow}+k^x_{i,\downarrow})^* & 2J_i^x p^x(i+x) & 0 \\
    0 & 2J_i^x p^x(i) & t-2J_i^x(k^x_{i,\uparrow}+k^x_{i,\downarrow}) & 0\\
    2J_i^x p^x(i+x) & 0 & 0 & t-2J_i^x(k^x_{i,\uparrow}+k^x_{i,\downarrow})
  \end{array}
  \right)
  \end{equation}

and
\begin{equation}
  \underline{\underline{V}}_{i}(k_y)=\left(
  \begin{array}{cccc}
    -2t\cos(k_y)+v_{\uparrow,\uparrow}(i) & v_{\uparrow,\downarrow}(i) & 0 & U_i f_0(i)+2J_i^y p^y(i) \\
    v_{\downarrow,\uparrow}(i) & -2t\cos(k_y)+v_{\downarrow,\downarrow}(i) & U_i f_0(i)+2J_i^y p^y(i) & 0 \\
    0 & U_i \Delta^*_i+2J_i^y [p^y(i)]^* &  2t\cos(k_y)-v_{\uparrow,\uparrow}(i) & v^*_{\uparrow,\downarrow}(i) \\
    U_i \Delta^*_i+2J_i^y [p^y(i)]^* & 0 & v^*_{\downarrow,\uparrow}(i) & 2t\cos(k_y)-v_{\downarrow,\downarrow}(i)
    \end{array}
  \right)
  \end{equation}

\end{widetext}

with the following abbreviations
\begin{eqnarray*}
  v_{\sigma,\sigma}(i)&=& \frac{U_i}{2}(n_i-\sigma m_i)+V^{loc}_i-\mu +h_{i,z} \sigma \\
  &-&J^x_{i-x}(n_{i-x}+\sigma m_{i-x}) -J^x_i (n_{i+x}+\sigma m_{i+x}) \\
  &-&2 J^y_i(n_i+\sigma m_i)+ 4J^y_i Re[(k^y_{i,\uparrow}+k^y_{i,\downarrow})e^{ik_y}]\\
  v_{\uparrow,\downarrow}(i)&=& -U_i \langle S^-_i\rangle -2J^x_{i-x} \langle S^-_{i-x}\rangle -2 J^x_{i}\langle S^-_{i+x}\rangle\\ &-& 4 J^y_i \langle S^-_i\rangle\\
  n_i &=& \sum_{\sigma}\langle n_{i,\sigma}\rangle \\
  m_i &=& \sum_{\sigma}\sigma \langle n_{i,\sigma}\rangle\,\,\, (\sigma=\pm 1) \\
\langle S^-_i\rangle   &=& \langle c^\dagger_{i,\downarrow} c_{i,\uparrow}\rangle = \frac{1}{N_y}\sum_{k_y} \langle c^\dagger_{i,\downarrow}(k_y) c_{i,\uparrow}(k_y)\rangle
\end{eqnarray*}
\begin{eqnarray*}
  k^x_{i,\sigma}&=& \langle c^\dagger_{i,\sigma} c_{i+x,\sigma}\rangle = \frac{1}{N_y}\sum_{k_y} \langle c^\dagger_{i,\sigma}(k_y) c_{i+x,\sigma}(k_y)\rangle\\
  k^y_{i,\sigma}&=& \langle c^\dagger_{i,\sigma} c_{i+y,\sigma}\rangle 
  = \frac{1}{N_y}\sum_{k_y} e^{-ik_y} \langle c^\dagger_{i,\sigma}(k_y) c_{i,\sigma}(k_y)\rangle \\
p^x(i)&=& \langle c_{i,\downarrow} c_{i+x,\uparrow}\rangle = \frac{1}{N_y}\sum_{k_y} \langle c_{i,\downarrow}(k_y) c_{i+x,\uparrow}(k_y)\rangle\\
p^y(i)&=& \langle c_{i,\downarrow} c_{i+y,\uparrow}\rangle e^{-ik_y}+\langle c_{i+y,\downarrow} c_{i,\uparrow}\rangle e^{ik_y}
\,.
\end{eqnarray*}

\section{}\label{apb}
In Sec.~\ref{sec:resF} we discuss the magnetic state of the F alone, in the correlated single band model, Eq.~\eqref{eigenvalueproblem}. Spiral magnetic solutions with the Ansatz
$\langle S^\pm_i \rangle = S_0 \exp(\pm i \qvec {\bf R_i})$
are obtained by factorizing
Eqs. (\ref{eq:hu}, \ref{eq:hint}) with respect to the operators
\begin{displaymath}
  S^{+(-)}_q = \sum_\kvec c^{\dagger}_{\kvec+\qvec \uparrow(\downarrow)} c^{}_{\kvec \downarrow(\uparrow)}\,.
\end{displaymath}  
Since the charge density for these solutions is constant Hartree terms
are neglected as they only shift the energy by a constant value.
For a given momentum $\qvec$ the resulting energy is given by
\begin{widetext}
\begin{align}
\begin{split}
E(\qvec) = \sum^N_{\kvec} \left\langle \left( \begin{array}{c} c^{\dagger}_{\kvec + \qvec \uparrow} c^{\dagger}_{\kvec \downarrow}  \end{array} \right)
\underline{\underline{H}} \left( \begin{array}{c} c^{}_{\kvec+\qvec \uparrow}  \\ c^{}_{\kvec \downarrow}  \end{array} \right) \right\rangle
+ 4JNS_0^2 (cos(q_x) + cos(q_y)) - 2JN(v_x^2 + v_y^2) + UNS_0^2,
\end{split}
\end{align}
where
\begin{align}
\underline{\underline{H}} = \left( {\begin{array}{cc} \varepsilon_{\kvec + \qvec} + \frac{2J}{t} \left[ v_x \cos(k_x + q_x) + v_y \cos(k_y+q_y)\right] & -\frac{4JS_0}{t} \left[\cos(q_x) + \cos(q_y)\right] - \frac{US_0}{t} \\ -\frac{4JS_0}{t} \left[\cos(q_x) + \cos(q_y)\right] - \frac{US_0}{t} & \varepsilon_\kvec + \frac{2J}{t} \left[ v_x \cos(k_x) + v_y \cos(k_y) \right] \end{array} } \right)\,.
\end{align}
\end{widetext}
The quantities
\begin{displaymath}
v_{x/y} = \frac{1}{N} \sum_{k \sigma} \cos(k_{x/y}) \langle n_{\kvec \sigma} \rangle\,.
\end{displaymath}
renormalize the kinetic energy via the magnetic interaction and have to be
determined self-consistently.

\end{document}